\documentstyle[12pt]{article}
\def\href#1#2{{#2}}
\begin{document}
\begin{titlepage}
\begin{flushright}
quant-ph/9503009 \\
THEP-95-2 \\
March 1995
\end{flushright}
\vspace{0.2cm}
\begin{center}
\LARGE
Standard Model plus Gravity \\
from Octonion Creators and Annihilators \\
\vspace{1cm}
\normalsize
Frank D. (Tony) Smith, Jr. \\
\footnotesize
e-mail: gt0109e@prism.gatech.edu \\
and fsmith@pinet.aip.org  \\
P. O. Box for snail-mail: \\
P. O. Box 430, Cartersville, Georgia 30120 USA \\
\href{http://www.gatech.edu/tsmith/home.html}{WWW
URL http://www.gatech.edu/tsmith/home.html} \\
\vspace{12pt}
School of Physics  \\
Georgia Institute of Technology \\
Atlanta, Georgia 30332 \\
\vspace{0.2cm}
\end{center}
\normalsize
\begin{abstract}
Octonion creation and annihilation operators are
used to construct the Standard Model plus Gravity.
The resulting phenomenological model is the
$D_{4}-D_{5}-E_{6}$ model described in
\href{http://xxx.lanl.gov/abs/hep-ph/9501252}{hep-ph/9501252}.
\end{abstract}
\vspace{0.2cm}
\normalsize
\footnoterule
\noindent
\footnotesize
\copyright 1995 Frank D. (Tony) Smith, Jr., Atlanta, Georgia USA
\normalsize
\end{titlepage}
\newpage
\setcounter{footnote}{0}
\setcounter{equation}{0}

\tableofcontents

\newpage

\section{Introduction.}

The purpose of this paper is to outline a way to
build a model of the Standard Model plus Gravity
from the Heisenberg algebra of fermion creators and
annihilators.
\vspace{12pt}

We want to require that the superposition space of
charged fermion creation operators be
represented by multiplication on a continuous unit
sphere in a division algebra.  That limits us to:
\vspace{12pt}

the complex numbers $\bf{C}$, with parallelizable $S^{1}$,
\vspace{12pt}

the quaternions $\bf{Q}$, with parallelizable $S^{3}$, and
\vspace{12pt}

the octonions $\bf{O}$, with parallelizable $S^{7}$.
\vspace{12pt}

We choose the octonions because they are big enough to
make a realistic physics model.
\vspace{12pt}

Octonions are described in Geoffrey Dixon's book \cite{DIX4}
and subsequent papers \cite{DIX5, DIX6, DIX7, DIX8}, and
in Ian Porteous's book \cite{POR}.  Essential mathematical
tools include the octonion X-product of Martin Cederwall \cite{CED1}
and the octonion XY-product of Geoffrey Dixon \cite{DIX8}.
\vspace{12pt}

The purpose of this paper is to build a physics model,
not to do mathematics, so I ignore mathematical details
and subtle points.  For them, see the references.
\vspace{12pt}

This paper is the result of discussions with Ioannis Raptis
and Sarah Flynn, and reading a preprint of Steve Selesnick
on fermion creation operators as fundamental to the
Quantum Net of David Finkelstein.  John Caputlu-Wilson has
discussed the role of propagator phase.  Igor Kulikov and
Tang Zhong have also discussed the paper, and Igor has
made it clear that I should not misspell Shilov as Silov.
\vspace{12pt}

\newpage

\section{Octonion Creators and Annihilators.}

Consider the octonions $\bf{O}$ and their unit sphere $S^{7}$.
\vspace{12pt}

Our starting point is the creation
operator $\alpha_{{\bf{O}}L}$ for
the first generation octonion fermion particles.
In the octonion case, the $L$ denotes only
the helicity of the neutrino, which is a Weyl fermion.
The other fermions are Dirac fermions, and can
exist in either helicity state $L$ or $R$.
\vspace{12pt}

If a basis for the octonions is
$\{ 1, e_{1}, e_{2}, e_{3}, e_{4}, e_{5}, e_{6}, e_{7}, \}$,
then the first generation fermion particles are
represented by:
\vspace{12pt}

\begin{equation}
\begin{array}{|c|c|} \hline
Octonion  & Fermion \: Particle \\
basis \: element & \\ \hline
1 & e-neutrino   \\ \hline
e_{1} & red \: up \: quark \\ \hline
e_{2} & green \: up \: quark \\ \hline
e_{6} & blue \: up \: quark  \\ \hline
e_{4} & electron \\ \hline
e_{3} & red \: down \: quark  \\ \hline
e_{5} & green \: down \: quark  \\ \hline
e_{7} & blue \: down \: quark  \\ \hline
\end{array}
\end{equation}

\vspace{12pt}

Graphically, represent the neutral left-handed Weyl
e-neutrino creation operator $\alpha_{\nu_{e}L}$ by
\vspace{12pt}

\begin{picture}(50,50)

\put(20,25){\circle*{3}}
\put(20,25){\vector(1,0){20}}
\put(10,18){$\alpha_{\nu_{e}L}$}

\end{picture}

\vspace{12pt}

Now, represent the charged left-handed and right-handed
Dirac electron-

\newpage

quark creation operators
$\alpha_{eqL}$ and $\alpha_{eqL}$ by vectors to a point
on the sphere $S^{7}$ (represented graphically by a circle):
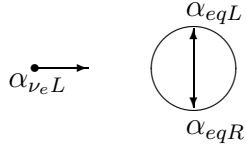
\vspace{12pt}

\begin{picture}(100,50)

\put(80,25){\circle{30}}
\put(20,25){\circle*{3}}
\put(80,25){\vector(0,1){15}}
\put(80,25){\vector(0,-1){15}}
\put(20,25){\vector(1,0){20}}
\put(77,45){$\alpha_{eqL}$}
\put(77,0){$\alpha_{eqR}$}
\put(10,18){$\alpha_{\nu_{e}L}$}

\end{picture}

\vspace{12pt}

Any superposition of charged fermion particle creation
operators $\alpha_{eqL}$ and $\alpha_{eqR}$
can be represented as a point on the sphere $S^{7}$
defined by their representative vectors.
The sphere $S^{7}$ should be thought of as being orthogonal
to the vector $\alpha_{\nu_{e}L}$.
\vspace{12pt}

We will represent the superposition of creation of e-neutrinos
(represented by a vector on a line)
and charged particles
(represented by vectors to a sphere $S^{7}$)
by letting the magnitude of the amplitude
$\mid \alpha_{\nu_{e}L} \mid$ of the e-neutrino creator vector
run from 0 to 1 and then determining the radius $r$ of the
sphere $S^{7}$ in octonion space $\bf{O}$ by

\begin{equation}
\mid \alpha_{\nu_{e}L} \mid^{2}  + r^{2} = 1
\end{equation}

\vspace{12pt}

We now have as representation space for the octonion creation
operators $S^{7} \times {\bf{R}}P^{1}$, where we have
parameterized  ${\bf{R}}P^{1}$ by the interval $[0,1)$
rather than the conventional $[0,\pi )$.
\vspace{12pt}

The octonion first-generation fermion annihilation
operator, or antiparticle creation operator, is
$\alpha^{\dagger}_{{\bf{O}}R}$.
\vspace{12pt}

Therefore, for the octonions, we have the
nilpotent Heisenberg algebra matrix:
\vspace{12pt}

\begin{equation}
\left(
\begin{array}{ccc}
0 & \alpha_{{\bf{O}}L} & \beta \\
& & \\
0 & 0 & \alpha^{\dagger}_{{\bf{O}}R} \\
& & \\
0 & 0 & 0
\end{array}
\right)
\end{equation}

\vspace{12pt}

How does this correspond to the $D_{4}-D_{5}-E_{6}$ model
described in
\href{http://xxx.lanl.gov/abs/hep-ph/9501252}{hep-ph/9501252}
\cite{SMI7}?
\vspace{12pt}

The octonion fermion creators and annihilators,
\vspace{12pt}

\begin{equation}
\left(
\begin{array}{ccc}
0 & \alpha_{{\bf{O}}L} & 0 \\
& & \\
0 & 0 & \alpha^{\dagger}_{{\bf{O}}R} \\
& & \\
0 & 0 & 0
\end{array}
\right)
\end{equation}

\vspace{12pt}

are both together represented in the $D_{4}-D_{5}-E_{6}$
model by the Shilov boundary of the bounded complex
homogeneous domain corresponding to the Hermitian
symmetric space $E_{6}/(D_{5} \times U(1))$.
\newline
(A good reference on Shilov boundaries is
Helgason \cite{HEL2}.)
\newline
The Shilov boundary is two copies of
$S^{7} \times {\bf{R}}P^{1}$.
The ${\bf{R}}P^{1}$ part represents the Weyl neutrino,
and the $S^{7}$ part represents the Dirac electron
and red, green, and blue up and down quarks.
\vspace{12pt}

The ${\bf{R}}P^{1}$ part is represented by $[0,1)$ in our
parameterization (or $[0,\pi)$ on the unit circle in the complex
plane in a more conventional one), and the $S^{7}$ part can
be represented by the unit sphere $S^{7}$ in the
octonions $\bf{O}$.
\vspace{12pt}

Also, mathematically, we can regard
\vspace{12pt}

\begin{equation}
S^{7} \times {\bf{R}}P^{1} = (S^{7} \times
{\bf{R}}P^{1})^{\dagger}
\end{equation}

\vspace{12pt}

Therefore, the creator-annihilator part of the
nilpotent Heisenberg $3 \times 3$ matrix can be
represented as:

\vspace{12pt}

\begin{equation}
\left(
\begin{array}{ccc}
0 & S^{7} \times {\bf{R}}P^{1} & 0 \\
& & \\
0 & 0 & S^{7} \times {\bf{R}}P^{1} \\
& & \\
0 & 0 & 0
\end{array}
\right)
\end{equation}

\vspace{12pt}

What about the $\beta$ part?
\vspace{12pt}

$\beta$ is given by the commutator

\begin{equation}
\beta = [S^{7} \times {\bf{R}}P^{1},
S^{7} \times {\bf{R}}P^{1}]
\end{equation}

\vspace{12pt}

Since ${\bf{R}}P^{1}$ is only the interval
$[0,1)$ in our parameterization
(or $[0,\pi)$ on the unit circle in the complex
plane in a more conventional one),
it is equivalent to a real number and
can therefore be absorbed into the real $R$
scalar field of the $3 \times 3$ matrices.
\newline
It commutes with everything and produces
no gauge bosons by its commutators.
\vspace{12pt}

{}From a physical point of view, we can say that
${\bf{R}}P^{1}$ represents the neutrino, which
has no charge and therefore does not interact with
or produce any gauge bosons by commutation.
\vspace{12pt}

Whichever point of view you prefer, the result
is that the full $3 \times 3$ nilpotent Heisenberg
matrix looks like:
\vspace{12pt}

\begin{equation}
\left(
\begin{array}{ccc}
0 & S^{7} & \beta \\
& & \\
0 & 0 & S^{7} \\
& & \\
0 & 0 & 0
\end{array}
\right)
\end{equation}

\vspace{12pt}

Therefore, $\beta$ is given by

\begin{equation}
\beta =  [S^{7},S^{7}]
\end{equation}

\vspace{12pt}

Unlike the parallelizable spheres $S^{1}$ and $S^{3}$
of the associative algebras $\bf{C}$ and $\bf{Q}$, the
7-spehre $S^{7}$ of the nonassociative octonions $\bf{O}$
does not close under commutator and does
not form a Lie algebra.
\vspace{12pt}

To deal with the situation, we need to use
Martin Cederwall's octonion X-product \cite{CED1} and
Geoffrey Dixon's XY-product \cite{DIX8}.
\vspace{12pt}

Martin Cederwall and his coworkers \cite{CED1}
have shown that $[S^{7},S^{7}]$ does form an algebra,
but not a Lie algegbra:
\vspace{12pt}

Consider a basis $\{ e_{iX} \}$ of the tangent space of
$S^{7}$ at the point $X$ on $S^{7}$.
Following Cederwall and Preitschopf \cite{CED1}, we have
\vspace{12pt}

\begin{equation}
[e_{iX},e_{jX}] = 2 T_{ijk}(X) e_{kX}
\end{equation}

\vspace{12pt}

Due to the nonassociativity of the octonions, the
"structure constants" $T_{ijk}(X)$ are not constant,
but vary with the point $X$ on $S^{7}$, producing torsion.
\vspace{12pt}

Effectively, each point of $S^{7}$ has its own
X-product algebra.
\vspace{12pt}

The X-product algebra takes care of the case of
$[e_{iX},e_{jX}]$ where both of the elements are in the
tangent space of the same point $X$ of $S^{7}$,
but since different points have
really different tangent spaces due to nonassociativity
of the octonions, it does not take care of the
case of $[e_{iX},e_{jY}]$ where $e_{iX}$ is an element
of the tangent space at $X$ and $e_{iY}$ is an element
of the tangent space at $Y$.
\vspace{12pt}

To take care of this case, we must use Geoffrey Dixon's
XY-product and "expand" $[S^{7},S^{7}]$ from $S^{7}$
to at least two copies of $S^{7}$ (one for the commutor algebra at
each of the points of the other one).  That is, if $\Join$
denotes a fibration "product":
\vspace{12pt}

\begin{equation}
[S^{7},S^{7}] \supset S^{7} \Join S^{7}
\end{equation}

\vspace{12pt}

We are still not quite through, because even though we
have used the XY-product to take care of the
case of $[e_{iX},e_{jY}]$ where $e_{iX}$ is an element
of the tangent space at $X$ and $e_{iY}$ is an element
of the tangent space at $Y$, we have not taken into account
that the octonion basis for the tangent spce at at $X$ may be
significantly different from the octonion basis for the
tangent space at $Y$.
\vspace{12pt}

The extra structure that must be "added" to $S^{7} \Join S^{7}$
to "transform" the tangent space at $X$ into
the tangent space at $Y$ is the automorphism group $G_{2}$ of
the octonions.  Unlike the cases of the associative algebras,
the action the automorphism group cannot be absorbed into
the products we have already used.  So, we see that the
Lie algebra of $[S^{7},S^{7}]$ is
\vspace{12pt}

\begin{equation}
[S^{7},S^{7}] = S^{7} \Join S^{7} \Join G_{2} = Spin(8)
\end{equation}

\vspace{12pt}

The fibrations represented by the $\Join$ are:

\begin{equation}
Spin(7) \rightarrow Spin(8) \rightarrow S^{7}
\end{equation}
and
\begin{equation}
G_{2} \rightarrow Spin(7) \rightarrow S^{7}
\end{equation}

\vspace{12pt}

Now, our octonionic version of the nilpotent
Heisenberg algebra looks like:
\vspace{12pt}

\begin{equation}
\left(
\begin{array}{ccc}
0 & S^{7} & Spin(8) \\
& & \\
0 & 0 & S^{7} \\
& & \\
0 & 0 & 0
\end{array}
\right)
\end{equation}

\vspace{12pt}

Here, $Spin(8)$ is the 28-dimensional adjoint representation
of $Spin(8)$.  Its 28 infinitesimal generators represent
28 gauge bosons acting on the fermions that we have
created, all as in the $D_{4}-D_{5}-E_{6}$ model.
\vspace{12pt}

The action of the $Spin(8)$ gauge bosons takes place
within the arena of the 8-dimensional vector representation
of $Spin(8)$, again as in the $D_{4}-D_{5}-E_{6}$ model.
\vspace{12pt}

We now have the picture of fermion creators and annihilators
forming gauge bosons, and all of them interacting in
accord with the $D_{4}-D_{5}-E_{6}$ model.
\vspace{12pt}

However, what about spacetime?
\vspace{12pt}

Since by triality (Porteous \cite{POR} describes triality)
the vector representation of $Spin(8)$ is isomorphic to
each of the half-spinor representations
that we use for fermion creators and annihilators,
we can form a vector representation version of the octonionic
nilpotent Heisenberg algebra.
\vspace{12pt}

\begin{equation}
\left(
\begin{array}{ccc}
0 & S^{7} & S^{7} \\
& & \\
0 & 0 & S^{7} \\
& & \\
0 & 0 & 0
\end{array}
\right)
\end{equation}

\vspace{12pt}

If we put back explicitly the factors of $\bf{R}P^{1}$
that we had merged into the real scalar field for
ease of calculation of the $S^{7}$ commutators,
we get:
\vspace{12pt}

\begin{equation}
\left(
\begin{array}{ccc}
0 & S^{7} \times {\bf{R}}P^{1} & S^{7} \times {\bf{R}}P^{1} \\
& & \\
0 & 0 & S^{7} \times {\bf{R}}P^{1} \\
& & \\
0 & 0 & 0
\end{array}
\right)
\end{equation}

\vspace{12pt}

The vector $Spin(8)$ spacetime part is
\vspace{12pt}

\begin{equation}
\left(
\begin{array}{ccc}
0 & 0 & S^{7} \times {\bf{R}}P^{1} \\
& & \\
0 & 0 & 0 \\
& & \\
0 & 0 & 0
\end{array}
\right)
\end{equation}

\vspace{12pt}

It is represented in the $D_{4}-D_{5}-E_{6}$ model by
the Shilov boundary of the bounded complex homogeneous domain
corresponding to the Hermitian symmetric space
$D_{5}/(D_{4} \times U(1))$.
\newline
The Shilov boundary is $S^{7} \times {\bf{R}}P^{1}$.
The ${\bf{R}}P^{1}$ part represents the time axis,
and the $S^{7}$ part represents a 7-dimensional
space.
\vspace{12pt}

NOW, we have reproduced the structure of the
$D_{4}-D_{5}-E_{6}$ model by starting from
octonion fermion creators and annihilators.
\vspace{12pt}

We can therefore incorporate herein by
reference all the phenomenological results
of the $D_{4}-D_{5}-E_{6}$ model as described in
\href{http://xxx.lanl.gov/abs/hep-ph/9501252}{hep-ph/9501252}
\cite{SMI7}.
\vspace{12pt}

\section{Complexified Octonions.}
\vspace{12pt}

Recall that the octonion fermion creators and annihilators
are of the form
\vspace{12pt}

\begin{equation}
\left(
\begin{array}{ccc}
0 & S^{7} \times {\bf{R}}P^{1} & 0 \\
& & \\
0 & 0 & S^{7} \times {\bf{R}}P^{1} \\
& & \\
0 & 0 & 0
\end{array}
\right)
\end{equation}

\vspace{12pt}

and that both of the entries $S^{7} \times {\bf{R}}P^{1}$
taken together are represented in the $D_{4}-D_{5}-E_{6}$ \
model by the Shilov boundary of the bounded complex
homogeneous domain corresponding to the Hermitian
symmetric space $E_{6}/(D_{5} \times U(1))$.
\vspace{12pt}

Also recall that the vector $Spin(8)$ spacetime part
\vspace{12pt}

\begin{equation}
\left(
\begin{array}{ccc}
0 & 0 & S^{7} \times {\bf{R}}P^{1} \\
& & \\
0 & 0 & 0 \\
& & \\
0 & 0 & 0
\end{array}
\right)
\end{equation}

\vspace{12pt}

is also represented in the $D_{4}-D_{5}-E_{6}$ model by
a Shilov boundary of a bounded complex homogeneous domain.
This entry $S^{7} \times {\bf{R}}P^{1}$
corresponds to the Hermitian symmetric space
$D_{5}/(D_{4} \times U(1))$.
\vspace{12pt}

What if, instead of representing the $3 \times 3$
nilpotent Heisenberg matrix structure by Shilov boundaries,
we represent them by the linearized tangent spaces
of the corresponding Hermitian symmetric spaces?
\vspace{12pt}

Then we would have:
\vspace{12pt}

\begin{equation}
\left(
\begin{array}{ccc}
0 & {\bf{C}} \otimes {\bf{O}} & {\bf{C}} \otimes {\bf{O}} \\
& & \\
0 & 0 & {\bf{C}} \otimes {\bf{O}} \\
& & \\
0 & 0 & 0
\end{array}
\right)
\end{equation}

\vspace{12pt}

Sarah Flynn uses such $3 \times 3$ matrix structures in
her work \cite{FLY}.
\newline
Note that complexified octonions ${\bf{C}} \otimes {\bf{O}}$
are not a division algebra.
\newline
That is because signature is indistinguishable in complex
spaces.
\newline
Therefore, both the octonions and the split octonions
are subspaces of  ${\bf{C}} \otimes {\bf{O}}$.
\newline
Since the split octonions contain nonzero null vectors,
the complexified octonions ${\bf{C}} \otimes {\bf{O}}$
may be a normed algebra, but they are not a division
algebra.
\newline
The only complex division algebra is the
complex numbers $\bf{C}$ themselves.
\vspace{12pt}

\section{Dimensional Reduction.}

Now, going back to the Shilov boundary uncomplexified
representations, recall that the vector $Spin(8)$ spacetime
is represented by
\vspace{12pt}

\begin{equation}
\left(
\begin{array}{ccc}
0 & 0 & S^{7} \times {\bf{R}}P^{1} \\
& & \\
0 & 0 & 0 \\
& & \\
0 & 0 & 0
\end{array}
\right)
\end{equation}

\vspace{12pt}

Here, the spacetime of the vector representation of
$Spin(8)$ is $S^{7} \times {\bf{R}}P^{1}$, which can
be represented by the octonions if ${\bf{R}}P^{1}$ is
the real axis and $S^{7}$ is the imaginary octonions.
\vspace{12pt}

How do we move a fermion created at one point to another point?
\vspace{12pt}

If we move a particle along a lightcone path,
how do we tell how "far" it has gone?
\vspace{12pt}

Following the approach of John Caputlu-Wilson \cite{CAP},
we should measure how much its propagator phase has advanced.
\vspace{12pt}

Since the phase advance may be greater than $2 \pi$,
the propagator phase should take values, not on the
unit circle, but on the infinite helical multivalued
covering space of the unit circle.
\vspace{12pt}

Recognizing that it may be difficult to do an experiment
that will distinguish phases $\theta$ greater than $2 \pi$
from phases $\theta - 2 \pi$, we will look at very short
paths such that the phase advance along the path is much
less than $2 \pi$.
\vspace{12pt}

Now that we have a way to tell how "long" is a lightcone
path segment, we can look at some paths.
\newline
Consider the following two lightcone paths $P1$ and $P2$,
each beginning at $X$ and ending at $Y$ and
each made up of two "short" lightcone segments:
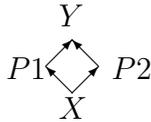
\vspace{12pt}

\begin{picture}(50,50)

\put(25,10){\vector(1,1){10}}
\put(25,10){\vector(-1,1){10}}
\put(35,20){\vector(-1,1){10}}
\put(15,20){\vector(1,1){10}}
\put(40,15){$P2$}
\put(0,15){$P1$}
\put(20,35){$Y$}
\put(20,0){$X$}

\end{picture}

\vspace{12pt}

Since the octonion spacetime $S^{7} \times {\bf{R}}P^{1}$
is nonassociative, it has
\newline
(as Martin Cederwall and his coworkers have shown \cite{CED1})
\newline
torsion.
\vspace{12pt}

Since it has torsion, the end-point $Y$ may not be
well-defined, and we may have the diagram:
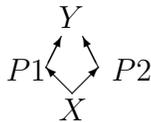
\vspace{12pt}

\begin{picture}(50,50)

\put(25,10){\vector(1,1){10}}
\put(25,10){\vector(-1,1){10}}
\put(35,20){\vector(-1,2){6}}
\put(15,20){\vector(1,2){6}}
\put(40,15){$P2$}
\put(0,15){$P1$}
\put(20,35){$Y$}
\put(20,0){$X$}

\end{picture}

\vspace{12pt}

Since we want paths and lightcones to be consistently
defined in the Minkowski vacuum spacetime
\newline
(before gravity has acted to effectively distort spacetime)
\newline
we must modify our octonionic spacetime so that it is
torsion-free at the Minkowski vacuum level.
\vspace{12pt}

How do we get rid of the torsion?
\vspace{12pt}

We must get rid of the nonassociativity.
\vspace{12pt}

To do that, reduce the octonionic spacetime
$S^{7} \times {\bf{R}}P^{1}$ to its
maximal associative subspace.
\vspace{12pt}

How do we determine the maximal associative subspace
of the octonionic spacetime $S^{7} \times {\bf{R}}P^{1}$?
\vspace{12pt}

Following Reese Harvey \cite{HAR}, define the
associative 3-form $\phi(x,y,z)$ for $x,y,x \in S^{7}$ by:
\vspace{12pt}

\begin{equation}
\phi(x,y,z) = <x,yz>
\end{equation}

\vspace{12pt}

where $<x,yz>$ is the octonion inner product
$Re(x \overline{yz})$ .
\vspace{12pt}

The associative form $\phi(x,y,z)$ is a calibration that
defines an associative submanifold of $S^{7}$.
\vspace{12pt}

When combined with the real axis part ${\bf{R}}P^{1}$ of
octonion spacetime, the associative submanifold of $S^{7}$
gives us a 4-dimensional quaternionic associative spacetime
submanifold of the type $S^{3} \times {\bf{R}}P^{1}$.
\vspace{12pt}

4-DIMENSIONAL QUATERNIONIC SPACETIME
$S^{3} \times {\bf{R}}P^{1}$
\newline
IS THE ASSOCIATIVE PHYSICAL SPACETIME.
\vspace{12pt}

This structure is the same as that of the $D_{4}-D_{5}-E_{6}$
model.  A detailed description of how dimensional reduction
works in the $D_{4}-D_{5}-E_{6}$ model, including its effects
on fermions and guage bosons, is given in
\href{http://xxx.lanl.gov/abs/hep-ph/9501252}{hep-ph/9501252}
\cite{SMI7}.
\vspace{12pt}

\newpage

\section{Spacetime and Internal Symmetries.}

Much of the material in this section is taken from the
book of Reese Harvey \cite{HAR}.  To the extent that this
section is good, he deserves credit.  To the extent that
this section is wrong or bad, it is not his fault that I
made mistakes using his book.
\vspace{12pt}

The 4-dimensional associative physical spacetime is determined
by the associative 3-form $\phi(x,y,z)$ on $S^{7}$ defined in
the previous section.
\vspace{12pt}

What happens to the rest of the original 8-dimensional
spacetime?
\vspace{12pt}

It is the orthogonal 4-dimensional space determined by the
coassociative 4-form $\psi(x,y,z,w)$ on $S^{7}$ defined
for $x,y,z,w$ in
$S^{7}$ as
\vspace{12pt}

\begin{equation}
\psi(x,y,z,w) = (1/2)(x,y({\overline{z}}w) - w({\overline{z}}y))
\end{equation}

\vspace{12pt}

That means that the original 8-dimensional spacetime
$S^{7} \times {\bf{R}}P^{1}$ is decomposed into an
associative physical spacetime
${\bf{\Phi}} = S^{3} \times {\bf{R}}P^{1}$ and
a coassociative internal space ${\bf{\Psi}}$ determined
by the coassociative 4-form $\psi(x,y,z,w)$ on $S^{7}$.
\vspace{12pt}

If the associative physical spacetime ${\bf{\Phi}}$ is taken to
be the real part and the coassociative internal space
${\bf{\Psi}}$ is taken to be the imaginary part of a
complex space ${\bf{\Phi}} + i{\bf{\Psi}}$, then the
full spacetime is transformed from a real 8-dimensional
space, locally ${\bf{R^{8}}}$, to a complex 4-dimensional
space, locally ${\bf{C^{4}}}$.
\vspace{12pt}

The gauge group $Spin(8)$ acting locally on ${\bf{R^{8}}}$
is then reduced to $U(4)$ acting locally on ${\bf{C^{4}}}$.
\vspace{12pt}

\subsection{Spacetime, Gravity, and Phase}

We now have the gauge group $U(4)$ acting on the
associative physical spacetime ${\bf{\Phi}}$ = $Re({\bf{C^{4}}})$.
\vspace{12pt}

Since $U(4)$ = $Spin(6) \times U(1)$, and $Spin(6)$ is the
compact version of the 15-dimensional conformal group,
we can now build a model of gravity by gauging the
conformal group $Spin(6)$ and use the $U(1)$ for the
phase of propagators in the associative physical
spacetime.
\vspace{12pt}

Note that only the 10-dimensional de Sitter gauge group
$Spin(5)$ subgroup of the $Spin(6)$ conformal group is
used to build gravity.
\vspace{12pt}

The other 5 degrees of freedom are 4 special conformal
transformations and 1 scale dilatation.  The 4 special
conformal transformations are gauge-fixed to pick
the $SU(2)$ symmetry-breaking direction of the Higgs
mechanism, and the scale dilatation is gauge-fixed to set
the Higgs mass scale.
\vspace{12pt}

For details, see
\href{http://xxx.lanl.gov/abs/hep-ph/9501252}{hep-ph/9501252}
\cite{SMI7} and \cite{SMI6}.
\vspace{12pt}

\subsection{Internal Space and Symmetries.}

Now we have:
\vspace{12pt}

associative physical spacetime
${\bf{\Phi}} = S^{3} \times {\bf{R}}P^{1} = Re({\bf{C^{4}}})$;
\vspace{12pt}

gravity from the conformal group $Spin(6)$;
\vspace{12pt}

Higgs symmetry breaking and mass scale from conformal $Spin(6)$;
and
\vspace{12pt}

propagator $U(1)$ phase.
\vspace{12pt}

We have not yet built anything from:
\vspace{12pt}

the coassociative imaginary space
${\bf{\Psi}}$ = $Im({\bf{C^{4}}})$ ; or
\vspace{12pt}

the part of the gauge group $Spin(8)$ that
is in the 12-dimensional coset space $Spin(8) / U(4)$ .
\vspace{12pt}

Let the coassociatve imaginary space
${\bf{\Psi}}$ = $Im({\bf{C^{4}}})$ be the internal symmetry
space on which the internal gauge groups act transitively.
\vspace{12pt}

That means that ${\bf{\Psi}}$ = $Im({\bf{C^{4}}})$ plays
a role similar to the internal symmetry spheres of
Kaluza-Klein models.
\vspace{12pt}

Let the part of the gauge group $Spin(8)$ that
is in the 12-dimensional coset space $Spin(8) / U(4)$
be the internal symmetry gauge groups.
\vspace{12pt}

A problem is presented here:
\newline
The coset space is just a coset space, with no group action.
\newline
How does it represent internal symmetry gauge groups?
\vspace{12pt}

The 12-dimensional coset space $Spin(8) / U(4)$
is the set of oriented complex structures
$Cpx^{+}(4)$ on ${\bf{R^{8}}}$, and
is also the Grassmannian $G_{\bf{R}}(2,{\bf{O}})$.
\vspace{12pt}

Each element of the Grassmannian $G_{\bf{R}}(2,{\bf{O}})$
can be represented by a simple unit vector in
$\bigwedge^{2}{\bf{O}}$.
\vspace{12pt}

Each simple unit vector in $\bigwedge^{2}{\bf{O}}$
determines a reflection, and all those reflections
generate the group $Spin(8)$.
\vspace{12pt}

Geometrically, what we have is that the
\newline
12-dimensional coset space $Spin(8) / U(4)$ can be represented
\newline
by 12 "positive" root vectors in the 4-dimensional root
\newline
vector space of the $D_{4}$ Lie algebra of $Spin(8)$,
\vspace{12pt}

while the 16-dimensional $U(4)$ subgroup of $Spin(8)$
can be represented by the 12 "negative" root vectors
plus the 4-dimensional Cartan subalgebra of the 4-dimensional
root vector space of the $D_{4}$ Lie algebra of $Spin(8)$.
\vspace{12pt}

Using quaternionic coordinates for the root vector space,
$$\{\pm 1, \pm i, \pm j, \pm k, (\pm 1 \pm i \pm j \pm k \}$$
are the 24 root vectors, and the 12-dimensional
coset space $Spin(8) / U(4)$ can be represented by
the 12 root vectors
$$\{+1, +i, +j, +k, (+ 1 \pm i \pm j \pm k)/2 \}$$
\vspace{12pt}

What internal symmetry gauge groups do the
12 coset space $Spin(8) / U(4)$ generators of $Spin(8)$ form?
\vspace{12pt}

Since the 12 coset space $Spin(8) / U(4)$ generators
can be represented by the quaternions
$$\{+1, +i, +j, +k, (+ 1 \pm i \pm j \pm k)/2 \}$$
and since they do not together form a simple Lie group,
consider what cartesian product of simple Lie groups might
be formed.
\vspace{12pt}

The 8 quaternions
$$\{(+ 1 \pm i \pm j \pm k)/2 \}$$
should form the Lie group $SU(3)$, with, for
example, $(+ 1 + i + j + k)/2 $ and $(+ 1 - i - j - k)/2$
as its Cartan sualgebra.
\vspace{12pt}

The 3 quaternions
$$\{+i, +j, +k \}$$\
should form the Lie group $SU(2)$, with, for
example, $ +j $ as its Cartan sualgebra.
\vspace{12pt}

The remaining quaternion
$$\{+1 \}$$\
should form the Lie group $U(1)$,
which is Abelian and equal to its Cartan sualgebra.
\vspace{12pt}

Therefore, in this model the 12-dimensional coset
space $Spin(8) / U(4)$ represents the internal symmetry
group of the Standard Model
$$SU(3) \times SU(2) \times U(1)$$.
\vspace{12pt}

The 4-dimensional internal symmetry space ${\bf{\Psi}}$ is the
representation space on which each of the internal
symmetry groups acts transitively.
\vspace{12pt}

The de Sitter $Spin(5)$ of the $U(4) = Spin(6) \times U(1)$
also acts transitively on the imaginary internal symmetry
space ${\bf{\Psi}}$.
\vspace{12pt}

Each of the 4 groups $Spin(5), SU(3), SU(2), U(1)$ act
transitively on the 4-dimensional internal symmetry
space ${\bf{\Psi}}$ with its own measure.
\vspace{12pt}

Effectively, each measure is determined by the way in
which the gauge bosons of each of the 4 forces "see"
the 4-dimensional internal symmetry space ${\bf{\Psi}}$.
\vspace{12pt}

The way each "sees" the space is determined by the
geometry of the 4-dimensional symmetric space ${\bf{\Psi}}_{force}$
on which each force acts transitively:
\vspace{12pt}

\begin{equation}
\begin{array}{|c|c|c|}
\hline
Gauge \: Group & Symmetric \: Space & {\bf{\Psi}}_{force}  \\
\hline
& &  \\
Spin(5) & Spin(5) \over Spin(4)  & S^4\\
& &  \\
SU(3) & SU(3) \over {SU(2) \times U(1)}
& {\bf C}P^2 \\
& & \\
SU(2) & SU(2) \over U(1)  & S^2 \times S^2 \\
& &  \\
U(1) & U(1)  & S^1 \times S^1 \times S^1
\times S^1 \\
& &  \\
\hline
\end{array}
\end{equation}

\vspace{12pt}

More about this is in
\newline
\href{http://www.gatech.edu/tsmith/See.html}
\newline
{WWW URL http://www.gatech.edu/tsmith/See.html} \cite{SMI6}.
\vspace{12pt}

The ratios of the respective measures are used to calculate
the relative force strength constants in this
$D_{4}-D_{5}-E_{6}$ model.  For detailed calculations of
force strengths (and also particle masses and K-M parameters),
see
\href{http://xxx.lanl.gov/abs/hep-ph/9501252}{hep-ph/9501252}
\cite{SMI7} and \cite{SMI6}.
\vspace{12pt}

Not only does the 10-dimensional de Sitter
\newline
$Spin(5)$ of the $U(4) = Spin(6) \times U(1)$ act on
\newline
the imaginary internal symmetry space, but
\newline
the (4+1)-dimensional conformal Higgs mechanism acts on
\newline
the internal symmetry space to give mass to
\newline
the $SU(2)$ weak bosons, and
\newline
the $U(1)$ propagator phase acts to give phases to the
gauge bosons.
\vspace{12pt}

NOW, we have constructed the $D_{4}-D_{5}-E_{6}$ model that
includes the Standard Model plus Gravity, all from the
beginning point of fermion creators and annihilators.
\vspace{12pt}

This construction of the model uses a continuous spacetime.
\newline
A future paper will deal with a discrete HyperDiamond lattice
generalized Feynman checkerboard version of the model.

\end{document}